\documentclass[11pt,a4paper]{article}
\usepackage[hyperref]{naaclhlt2018}
\usepackage{times}
\usepackage{latexsym}

\aclfinalcopy
\usepackage{times}
\usepackage{color}
\usepackage{amsfonts}
\usepackage{graphicx}
\usepackage{amsmath,amsfonts,amssymb,bm}
\usepackage{booktabs} 
\usepackage{array}
\usepackage{boxedminipage}
\usepackage{paralist}
\usepackage{url}
\usepackage{diagbox}
\usepackage{enumitem}
\usepackage{paralist}
\definecolor{darkspringgreen}{rgb}{0.05, 0.5, 0.06}
\usepackage{arydshln}

\usepackage{microtype}      % microtypography
\usepackage{comment}        % bulk comment
\usepackage{amsopn}
\usepackage{xcolor}
\usepackage{multirow}

\def\emnlppaperid{1301}

\newcolumntype{L}[1]{>{\centering\let\newline\\\arraybackslash\hspace{0pt}}m{#1}}
\newcolumntype{C}[1]{>{\centering\let\newline\\\arraybackslash\hspace{0pt}}m{#1}}
\newcolumntype{R}[1]{>{\centering\let\newline\\\arraybackslash\hspace{0pt}}m{#1}}

\title{Scientific Relation Extraction with Selectively Incorporated Concept Embeddings}
\author{
  Yi Luan\quad Mari Ostendorf \quad Hannaneh Hajishirzi \\
University of Washington\\
   {\{luanyi, ostendor, hannaneh\}@uw.edu}
}

\date{}

\begin{document}

\maketitle

\begin{abstract}

This paper describes our submission for the
SemEval 2018
Task 7 shared task on semantic relation extraction and classification in scientific papers. 
We extend the end-to-end relation extraction model of \cite{miwa2016end}  with enhancements such as a
character-level encoding attention mechanism on selecting pretrained concept candidate embeddings. Our official
submission ranked the second in relation classification task (Subtask 1.1 and Subtask 2 Senerio 2),
and the first in the relation extraction task (Subtask 2 Scenario 1).
%Scientific keyphrase extraction is the task of We present a CRF-LSTM neural approach for identifying keywords in scientific literature, in terms of Task, Process and Material. 
%In order to improve the performance of neural network based models with limited annotated training data and large unlabeled data, we explore different ways to use unlabeled data. We investigate the performance of word embedding trained on Semantic Scholar Computer Science data and Wikipedia data on different fields of papers (Computer Science, Physics and Material Science). The best model on different sub-fields is selected to do semi-supervised learning. Based on CRF-LSTM structure, we propose a way of reducing the effect of low confidence tokens, which gives state-of-the-art performance over all previous methods.

\end{abstract}

\section{Task Overview}

The SemEval 2018 Task 7 Shared Task \cite{SemEval2018Task7} focuses on the task of recognizing the semantic relation that holds between scientific concepts. The  task involves semantic relation extraction and classification into six categories specific to scientific literature: \textsc{usage}, \textsc{result}, \textsc{model-feature}, \textsc{part\_whole}, \textsc{topic}, \textsc{compare}. Two types of tasks are proposed:
1) identifying pairs of entities that are instances of any of the six semantic relations (extraction task), and
2) classifying instances into one of the specific relation types (classification task). 

Consider  the  following  input sentence:
``[\textit{Unsupervised training}] is first used to train a [\textit{phone n-gram model}] for a particular domain." 
Given the concept pair [\textit{Unsupervised training}] and [\textit{phone n-gram model}], the relation extraction task is to identify whether there is a relation between the concepts, while the the relation classification task is to identity the relation as \textsc{usage}. Relation directionality is not taken into account for the evaluation of the extraction task. Directionality is taken into account when relevant for the classification task (5 out of the 6 semantic relations are asymmetrical). We will use this example throughout the paper to illustrate various parts of our system.

The SemEval 2018 Task 7 dataset contains  350 abstracts from the ACL Anthology for training and validation, and 150 abstracts for testing each subtask. Since the scale of the data is small for supervised training of neural systems, we introduce several strategies to leverage a large quantity  of  unlabeled  scientific  articles. In addition to initializing a neural system with pre-trained word embeddings, as in \cite{luan2017scienceie}, we also try to incorporate embeddings of concepts 
%which are usually entities 
that span multiple words. 
In neural models such as \cite{miwa2016end}, phrases are often represented by an average (or weighted average) of the token's sequential LSTM representation. The intuition behind explicit modeling of multi-word concept embeddings is that the concept use may be different from that of its individual words.
%a concept is  less ambiguous than single words. 
Due to the size of the dataset and the nature of scientific literature, a large number of the scientific terms in the test set have never appeared in the training set, so supervised learning of the phrase embeddings is not feasible. Therefore, we  pre-trained scientific term embeddings on a large scientific corpus and provide a strategy to selectively incorporate the pre-trained embeddings into the relation extraction system.

\section{System Description}
\label{sec:IOB}

\subsection{Neural Architecture Model}
\label{sec:NN}
\begin{figure*}[t]
\centering
\includegraphics[width=16cm]{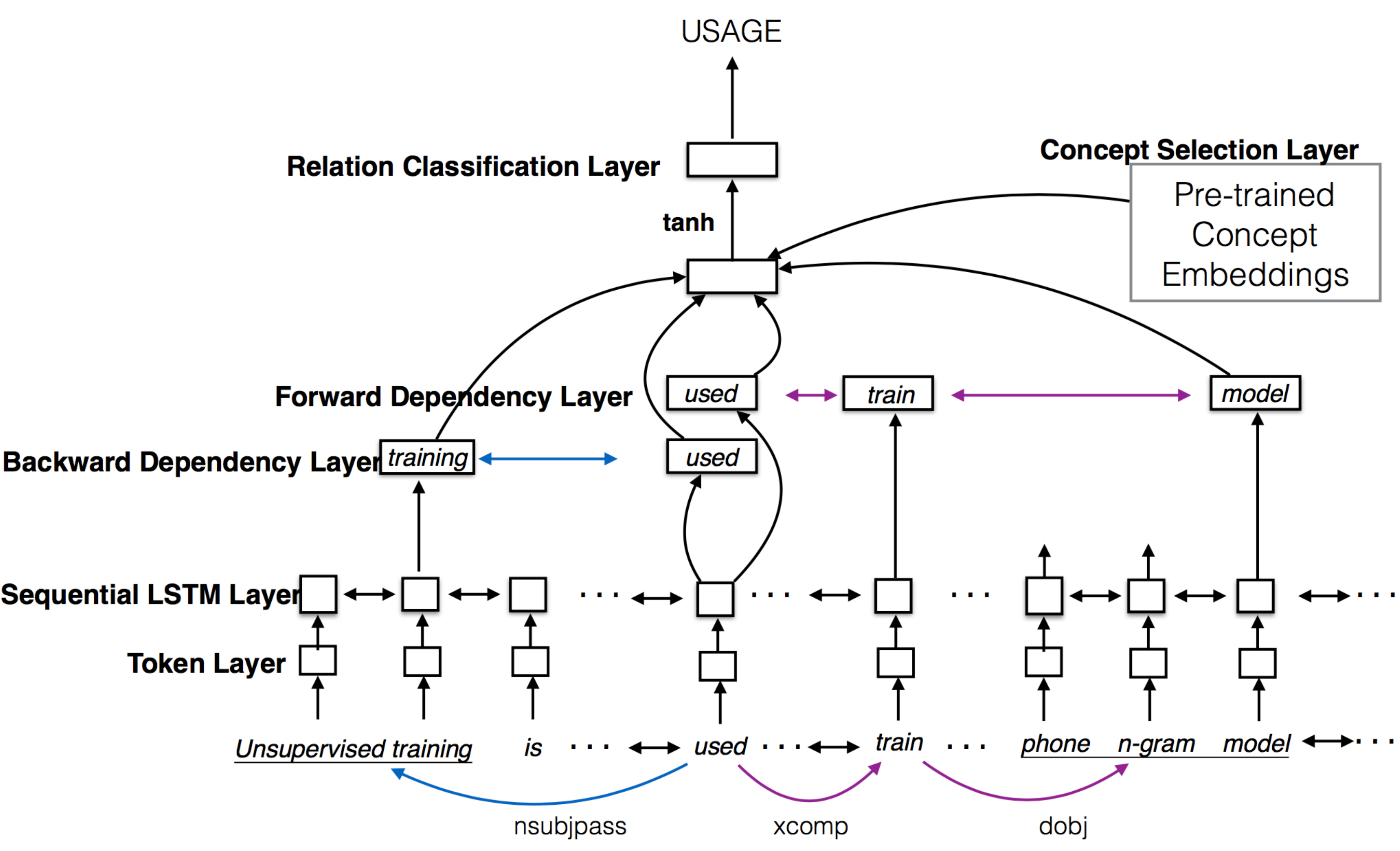}           
\caption{Neural relation extraction model with bidirectional sequential
and dependency path LSTMs.}
\label{fig:graph}
\end{figure*}

Our system is an extension of  \cite{luan2017scienceie} and \cite{miwa2016end} with LSTM RNNs that represent
both word sequences and dependency tree
structures, and perform 
relation extraction between concepts on top of these RNNs.
 %%MO: add cite for CoreNLP
As illustrated in Figure~\ref{fig:graph}, it is composed of a 5 types of layers in a hierarchical neural model to encode context information. The first two layers (token, token LSTM) use the neural modeling framework in \cite{luan2017scienceie}. The forward and backward dependency layers and the relation classification layer are based on \cite{miwa2016end}. The concept selection layer is novel, to the best of our knowledge.
The different layers are described in more detail below.

\begin{comment}
%\begin{enumerate}[topsep=2pt, leftmargin=0pt, itemsep=0pt]
    (1) The \textbf{Token Layer} concatenates   three components for each token: a bi-directional character-based embedding, a word embedding, and an embedding associated with orthographic and part-of-speech features. (2) The \textbf{Sequential LSTM layer} uses a bidirectional LSTM to incorporate contextual cues from surrounding tokens to derive intermediate token embeddings.    
     (3) \textbf{Forward \& Backward Dependency Layer} is bi-directional LSTM built on the shortest dependency path between the concept pairs.   
     (4) In order to incorporate the information of pre-trained concept embedding, \textbf{Concept Embedding Selection Layer} uses attention mechanism to select the most relevant concept candidate as additional feature.
     (5) \textbf{Relation Classification Layer} is a feed-forward structure which projects  feature from all other layers down to a lower dimension and then make relation predictions.
\end{comment}

\paragraph{Token Layer.}
The token layer concatenates three types of vector space embeddings.
%\paragraph{Word Embedding.} 
{\em Word embeddings} are learned for words from a fixed vocabulary (plus the unknown word token),
%are mapped to a vector space, 
initialized using Word2vec pre-training with large scholarly corpora. 
%
%\paragraph{Character-Based Embedding.}
The {\em character-based embedding} for a token is derived from its characters as the concatenation
of  forward  and  backward  representations  from
a bidirectional LSTM. The
character look-up table is initialized at random. The advantage of building a character-based embedding layer is that it can handle out-of-vocabulary words and equations, which are frequent in this data, all of which are mapped to ``UNK" tokens in the Word Embedding Layer. 
%
%\paragraph{Word Embedding.} 
{\em Word embeddings} are learned for words from a fixed vocabulary (plus the unknown word token),
%are mapped to a vector space, 
initialized using Word2vec pre-training with large scholarly corpora. 
%
%\paragraph{Feature Embedding.}
A {\em feature embedding} is learned as a mapping from features associated with capitalization (all capital, first capital, all lower, any capital but first letter) and part-of-speech tags. The embeddings are randomly initialized and trained jointly with other parameters
during supervised training.
%\mo{To YL: Please fix if you learn these in unsupervised training. I added this since I thought that's what you did and it's not clear from the text.}

\paragraph{Token LSTM Layer}
%%MO: since you are also using a bidirectional LSTM at the character level and don't give this much detail, it's not needed here
%Given a sequence of inputs $(x_{1},x_{2},\dots,x_{n})$ containing %$n$ tokens, each represented as a
%$d$-dimensional vector, an LSTM computes a representation $\bm{\overrightarrow{h}}_t$ of the left context of the sentence at every word $t$. Here, 
We apply a bidirectional LSTM at the token level taking the concatenated character-word-feature embedding as input. An LSTM hidden state generated in this layer is denoted as $h^S$.
%since right context $\bm{\overleftarrow{h}}_t$ as well  should add useful information. In order to obtain $\bm{\overleftarrow{h}}_t$, we use a second LSTM that reads in input word sequence in reverse order.  

%$\bm{h}_t = [\bm{\overrightarrow{h}}_t;\bm{\overleftarrow{h}}_t]$. Here we use the same implementation as \cite{lample2016neural} and the input vector for each word is the concatenation of Character Embedding Layer, Word Embedding Layer and Feature Layer.

\paragraph{Forward \& Backward Dependency Layers} 
Given the concept pair $(C_l,C_r)$, the Forward Dependency Layer (generating $h^F$) traces from the closest common ancestor $w_a$ (for example the word ``\textit{used}" in Fig.~\ref{fig:graph}) to the headword $w_j$ (word ``\textit{model}") of the right target  concept $C_r$ ( ``\textit{phone n-gram model}"). The Backward Dependency Layer (generating $h^B$) traces from the ancestor to the headword $w_i$ of the left concept $C_l$. We map the dependency relation into vector space and concatenate the resulting embedding to
the embedding ($h^S$) of the headword of the concepts $C_l$ or $C_r$ for the backward and forward dependency layers, respectively. We concatenate the resulting bi-directional LSTM vector for the headwords together with the common ancestor in both Forward \& Backward Dependency Layer as  input to Relation Classification Layer $h^{DP}=[\overleftarrow{h_{w_i}^B};\overrightarrow{h_{w_i}^B},\overleftarrow{h_{w_j}^F};\overrightarrow{h_{w_j}^F};\overleftarrow{h_{w_a}^B};\overrightarrow{h_{w_a}^B};\overleftarrow{h_{w_a}^F};\overrightarrow{h^F_{w_a}}]$ .

\paragraph{Concept Selection Layer}  
The concepts in the task are mostly phrases rather than single words, in the SemEval Task 7. We therefore seek ways to obtain prior knowledge for those terms. 
We train a scientific concept extraction model using the state-of-the-art scientific neural tagging technique in \cite{luan2017scienceie}, given the scientific concept annotation in the SemEval 2018 Task7 training data. We were able to achieve 79.8\% F1 score (span level) to identify the scientific concepts. We then use the model to extract all scientific concepts in the ACL anthology and AI2 dataset (refer to Sec.~\ref{sec:exp_setup}). We  keep all the concepts that occur more than 10 times in the whole corpus, which results in around 15k concepts. We treat each of the 15k concepts as an individual token and retrain word2vec embeddings $v_k$ together with all other single words. At training time, given a scientific concept pair $(C_l,C_r)$, we search through the 15k concepts to get all the concept candidates that have n-gram string match with $C_l$ and $C_r$ respectively (n is from 1 to the length of the target concept $C$). For example, for the concept  \textit{phone n-gram model}, the candidate concepts we get are \{\textit{phone n-gram}, \textit{n-gram model}, \textit{n-gram}, \textit{model}, \textit{phone}\}. 
Since there may exist cases where  no match could be found in the 15k concepts, we introduce a null vector $v_{\varnothing}$. $v_{\varnothing}$ is learned with other neural network parameters. Assume there are K concept candidates in the candidate list, we denote the embeddings for the concept candidates to be $V = \{v_1 \dots v_K, v_\varnothing \}$.
The attention weights are calculated by $\alpha_{lk} \propto \exp(h_{C_l}^S W_{ATT} v_k)$, where $v_k\in V$. $h_{C_l}^S$ is the concatenation of bidirectional LSTM hidden states of the first and last word in $C_l$.\footnote{We also tried using the
weighted average of all LSTM word embeddings in the span to calculate $h_{C_l}^S$; this yields a slightly worse result.} $W_{ATT}$ is a parameter matrix for the bilinear score for $h_{C_l}^S$ and $v_k$. The final concept embedding $v_{C_l}$ is $v_{C_l} = \sum_{v_k \in V} \alpha_{lk} v_k$. For a target concept C, if exact match exists in the 15K concepts, we set the pre-trained concept embedding to be $v_{C_l}$. We concatenate the resulting embedding for both  concepts  in the concept pair as input to the final classification layer ($v_C = [v_{C_l};v_{C_r}]$).

\paragraph{Relation Classification Layer}  
We concatenate the output of Forward \& Backward Dependency Layer $h^{DP}$ and Concept  Embedding  Selection Layer $v_C$ as input to Relation Classification Layer. Besides, we also introduce a distance feature between the two concepts which indicates how many other concepts there are in between the target concept pairs. We concatenate the distance embedding with all the other features. The concatenated features are then projected down to a lower dimension through $tanh$ function and  make the final prediction through a $softmax$ function.

% \begin{table*}
%   \centering
%  {\footnotesize
%   \begin{tabular}{llll}
%     \toprule
%   Span Level  & Classification (dev) & Classification (test)  & Identification \\
%   \midrule
%     Gupta et.al.(unsupervised)& - & 9.8 & 6.4\\
%     Tsai et.al. (unsupervised) & - & 11.9 & 8.0\\  \hdashline
%         \textsc{MultiTask} & 45.5 & - & -\\
%     Best Non-Neural SemEval$^+$ & - & 38 & 51 \\
%     Best Neural SemEval$^+$ & - & 44 & 56\\
%     \sys(supervised) & 48.1 & 40.2 & 52.1\\
%     \sys(semi)  & 51.9 & 45.3 & 56.9\\ \hdashline
%     \sys(semi)$^*$ & \textbf{52.1} & \textbf{46.6} & \textbf{57.6}\\
%     \bottomrule
%   \end{tabular}}
%   \caption{ \small{Overall span-level F1 results for keyphrase identification (SemEval Subtask A) and classification (SemEval Subtask B). $^*$ indicates tranductive setting. $^+$ indicates not documented as either transductive or inductive. - indicates score not reported or not applied.}}
%   \label{tab:best_span}
% \end{table*}

\section{Experimental Setup}
\label{sec:exp_setup}

\vspace{-.2cm}
\paragraph{External Data} We use two external resources for
pretraining  word embeddings: 
i) the Semantic Scholar Corpus,\footnote{http://labs.semanticscholar.org/corpus/}  a  collection  of over 20 million research  papers from which we extract a subset of 110k abstracts of publications in the artificial intelligence area; and ii) the ACL Anothology Reference Corpus, which contains 22k full papers published in the ACL Anothology~\cite{bird2008acl}.

\vspace{-.2cm}

\paragraph{Baseline} We 
%aim at analyzing the effect of the Concept Selection Layer, and 
compare our model with 
a  baseline that removes the Concept Selection Layer and replaces it with a weighted sum (using attention) of hidden states (from the Sequential LSTM Layer) for all words in a concept. 
%%MO: we need to explain the attention part

\vspace{-.2cm}

% \vspace{-.2cm}
\paragraph{Implementation details} %Training details are as follows:
% Pre-trained word2vec embeddings are tuned over dimensions $\{150,200,250,300\}$. Other parameters are randomly initialized. 
% The hidden dimension of forward and backward token-level embeddings is tuned over $\{50,100,150,200\}$ and over $\{10,25, 50\}$ for 
% forward and backward character embeddings.
% The dropout rate is fixed as 0.5
% We tried both Stochastic Gradient Descent (SGD) with the learning rate from $\{0.01,0.05, 0.1\}$ and Adam \cite{kingma2014adam}. 

All parameters are tuned based on dev set performance; the best parameters are selected and used for final evaluation. For all experiments, we explore tuning with two different evaluation metrics: macro-F1 score and micro-F1 score.\footnote{The official evaluation is macro-F1, but since the number of instances in each class is highly unbalanced, the observed macro-F1 scores were unstable. We therefore introduce micro-F1 score for tuning and evaluation as well.}
We keep the pre-trained concept embedding fixed as additional input feature. The word embedding dimension is 250; the LSTM hidden dimension is 100 (for both sequential and dependency layer); the character-level hidden dimension is 25; and the optimization algorithm is SGD with a learning rate of 0.05. For Subtask 2, since 5 out of 6 relation types have directionality, we add relation label ``\_REVERSE" to all the 5 directional relations together with a ``NONE" type, which result in 12 labels in total. For each epoch, we also randomly filter out some ``NONE" samples with  probability  $p$ during training, since the ``NONE" type relation dominates the training set and would bias the model towards predicting ``NONE" types. We tune $p$ according to dev set, and use $p=0.4$ for the final evaluation.

\begin{table}
  \centering
 {\footnotesize
  \begin{tabular}{l|lll|lll}
    \toprule
    & \multicolumn{3}{c}{Macro} & \multicolumn{3}{c}{Micro}  \\
     Model & P & R & F1 & P & R & F1  \\    
    \midrule
    Our system & 49.4 & 36.7 & \textbf{42.1} & 46.2 & 42.2 & \textbf{44.1}\\
     \midrule
     -DepFeat & 38.2& 39.6 & 39.0 & 45.2 & 41.9 & 43.0\\
     -DistFeat & 43.4 & 37.8 & 40.4 & 38.7 & 47.8 & 42.7\\
    -DepLSTM & 51.5 & 30.0 & 37.9 & 48.6 & 32.6 & 39.0\\
    -Concept & 36.2 & 41.8 & 38.8 & 37.6 & 46.5 & 41.6\\
    \midrule
    Baseline & 40.9& 32.5 & 36.2 & 41.9 & 38.0 & 39.9\\
    \bottomrule
  \end{tabular}}
  \caption{ Ablation study showing the impact of neural network configurations on system performance on the dev set for the relation classification task (Subtask 2, senerio 2).  
%  -DepFeat removes the input dependency relation embeddings from the Backward \& Forward Dependency Layer; -DistFeat omits the distance feature from the Relation Classification Layer;  -DepLSTM removes the Backward \& Forward Dependency Layers entirely (using the LSTM embeddings in the weighted token average); and -Concept omits the Concept Selection Layer. 
  -DepFeat removes the input dependency relation embeddings from the Backward \& Forward Dependency Layers. -DistFeat and -Concept omit the distance and concept selection features, respectively, from the final classification layer.  -DepLSTM removes the Backward \& Forward Dependency Layers entirely (using the LSTM embeddings in the weighted token average). 
  }
  \label{tab:ablation}
\end{table}

\section{Experimental Results}
\label{sec:exp_result}
\paragraph{Ablation Study}
Table \ref{tab:ablation} provides the results of an ablation study on the dev set showing the impact of removing different components of our system. Looking at micro F1 scores, dependency path information is very important (performance dropped 11.5\% without it), and the Concept Selection Layer is also important as it gives 2.5 absolute improvement. The Dependency relation feature and the distance feature also show 1-2 points gain. It is worth noticing that removing the Concept Layer (-Concept) does better than replacing it with the weighted sequential LSTM sum (Baseline). 
With the small amount of training data, it is difficult for the baseline system to learn a good transformation from word to phrase. 
%%MO: I think this is saying the same thing
%We observe that the phrase feature in baseline decrease the performance compared with removing the attention module (-Concept results in Table \ref{tab:ablation}), the reason may be due to the increased model complexity.

\paragraph{Competition Result}
The results of our system is in Table \ref{tab:result}. We submit two sets of results, one tuned with micro F1 and the other with macro F1. It turns out that even though the official evaluation metric is macro F1 score, our model tuned by micro F1 gets better results in the final competition. In Subtask 1.1 and Subtask 2 scenario 2, we were the second place team with F1 score of 78.9\% and 39.1\% respectively. We were the first place in  Subtask 2 scenario 1 with 50.0\% F1.

\begin{table}
  \centering
 {\footnotesize
  \begin{tabular}{l|lll}
    \toprule
     Model & T1.1 & T2-E & T2-C  \\    
    \midrule
   Our system (Micro) & 78.9 & \textbf{50.0} & 39.1 \\
   Our system (Macro) & 78.4 & 49.3 & 37.0 \\
   \midrule
    Team-1  & \textbf{81.7} & 48.8 & \textbf{49.3} \\
    Team-2 & 76.7 & 37.4 & 33.6 \\
%    Bf3R & - & 33.4 & 20.3 \\
%    VSP & - & 35.4 & 18.5 \\
    \bottomrule
  \end{tabular}}
  \caption{Competition result for the top 3 teams. The official evaluation metric is macro F1 score. T1.1 means Subtask 1.1, T2-E means Subtask 2 senerio 1 (extraction task), T2-C means Subtask 2 senerio 2 (classification task).} 
  \label{tab:result}
\end{table}

\section{Related Work}
There has been growing interest in research on automatic methods to help researchers search and extract information from scientific literature. Past research has addressed citation sentiment~\cite{athar2012detection,athar2012context}, citation networks~\cite{kas2011structures,GBOR16.870, sim2012discovering,do2013extracting,jaidka2014computational}, summarization~\cite{abu2011coherent} and some analysis of research community~\cite{vogel2012he,anderson2012towards}. However, due to scarce hand-annotated data resources, previous work on information extraction (IE) for scientific literature is very limited.  Most previous work focuses on unsupervised methods for extracting scientific terms such as bootstrapping \citet{gupta2011analyzing, tsai2013concept}, or extracting relations \cite{gabor2016unsupervised}.   \citet{luan2017scienceie, augenstein2017multi, luan2015efficient,luan2014semi, luan2016multiplicative} applied  semi-supervised learning and multi-task learning to neural based models to leverage large unannotated scholarly datasets for a scientific term extraction task \cite{augenstein2017multi}. 

Although not much supervised relation extraction work has been done on scientific literature,  neural network techniqueshave obtained 
the state of the art  for general domain relation extraction. Both convolutional \cite{santos2015classifying} and RNN-based architectures \cite{xu2016improved, miwa2016end, peng2017cross, quirk2016distant,luan2017multi} have been successfully applied to the task and significantly improve performance.

%Neural tagging models have been recently introduced to tagging problems such as NER. For example,  \citet{collobert2011natural} use a CNN over a sequence of word embeddings and apply a CRF layer on top.  \citet{huang2015bidirectional} use hand-crafted features with LSTMs to improve performance. There is currently great interest in using character-based embeddings in neural models.  \cite{chiu2015named,lample2016neural,ballesteros2015improved,ma2016end}. Our approach also takes advantage of neural tagging models and character-based embeddings for IE in scientific articles. 

% \rev{ Previous work on semi-supervised learning for neural models has mainly focused on transfer learning \cite{dai2015semi,luan2014semi,harsham2015driver} or initializing the model with pre-trained word embeddings \cite{mikolov2013efficient,pennington2014glove,levy2014dependency,luan2016lstm,luan2015efficient,luan2016multiplicative}. In our work, we use pre-training but also use more powerful methods including graph-based semi-supervision \cite{subramanya2011semi,liu2013graph,liu2015acoustic,liu2016graph,liu2016novel} and a method for leveraging partially labeled data \cite{kim2015weakly}. We show that the combination of these techniques gives better results than any one alone.}

\section{Conclusion}
This paper describes the system of the UWNLP team submitted to SemEval 2018 Task 7. We extend state-of-the-art neural models for information extraction by proposing a Concept Selection module which can leverage the semantic information of concepts pre-trained from a large scholarly dataset. Our system ranked second in the
relation classification task (subtask 1.1 and
subtask 2 senerio 2), and first in the relation extraction task (subtask 2 scenario 1).
\subsection*{Acknowledgments}
This research was supported by the NSF (IIS 1616112), Allen Distinguished Investigator Award, and gifts from Allen Institute of AI, Google, Amazon, Samsung, and Bloomberg. We thank the anonymous reviewers for their helpful comments
% This research was supported by the NSF (IIS 1616112), Allen Institute for AI (66-9175), Allen Distinguished Investigator Award, and gifts from Google, Samsung, and Bloomberg. We thank the anonymous reviewers for their helpful comments.

\bibliography{references}
\bibliographystyle{emnlp_natbib}

\end{document}